# ARTIFICIAL INTELLIGENCE IN CYBERSECURITY: BUILDING RESILIENT CYBER DIPLOMACY FRAMEWORKS

SUBMITTED TO

DR. DONOVAN CHAU

BY

MICHAEL STOLTZ

FOR

INR 5365

SPRING 2024

In a time of rapid technological progress, diplomacy is undergoing substantial changes. The integration of automation and artificial intelligence (AI) into diplomatic strategies, especially in cybersecurity, is becoming a key component of international relations.[1] This paper discusses the importance of automation and AI in enhancing United States (U.S.) cyber diplomacy, arguing that these technologies are not just supportive but transformative, managing the complexities and pace of modern diplomacy through improved decision-making, efficiency, and security measures.[2]

Cyber diplomacy, the practice of managing a nation's strategic interests in the digital domain, has become crucial as global interconnectivity increases.[3] In this context, the ability of automation and AI to process large volumes of data quickly and accurately is essential for responding to the dynamic threats and opportunities of the digital domain and cyberspace.[4] This paper emphasizes the importance of strategically incorporating these technologies into cyber diplomatic initiatives to maintain the U.S.'s competitive edge, secure national interests, and promote effective international cooperation.[5]

By adopting cutting-edge technology in diplomatic strategies, the U.S. can overcome previous limitations such as slow response times to cyber threats and labor-intensive analysis of international cyber policies.⁶ This introduction prepares for a deeper exploration of the transformative impact of automation and AI on U.S. cyber diplomacy, further discussing how these technologies enhance operational capabilities and strategic outcomes in subsequent sections.

---

[1] Amel Attatfa, Karen Renaud, and Stefano De Paoli, "Cyber Diplomacy: A Systematic Literature Review," *Procedia Computer Science* 176 (2020): 60-69, 61, https://doi.org/10.1016/j.procs.2020.08.007

[2] Alexandra-Cristina Dinu, "Cyber Diplomacy and Artificial Intelligence: Opportunities and Challenges," in *Proceedings of the International Conference on Cybersecurity and Cybercrime - 2023*, (Romania: Romanian Association for Information Security Assurance, 2023), 86-93, 89

[3] Simon Handler, "The 5×5—the Future of Cyber Diplomacy," *Atlantic Council*, September 29, 2021, 2, https://www.atlanticcouncil.org/commentary/the-5x5-the-future-of-cyber-diplomacy/.

[4] Alexandra-Cristina Dinu, "Cyber Diplomacy and Artificial Intelligence: Opportunities and Challenges," in *Proceedings of the International Conference on Cybersecurity and Cybercrime - 2023*, (Romania: Romanian Association for Information Security Assurance, 2023), 86-93, 89

[5] Simon Handler, "The 5×5—the Future of Cyber Diplomacy," *Atlantic Council*, September 29, 2021, 2, https://www.atlanticcouncil.org/commentary/the-5x5-the-future-of-cyber-diplomacy/.

Definitions of Key Concepts

Cyber diplomacy refers to managing a country's international relations and protecting its national interests within the digital domain, vital for advancing national interests, maintaining cybersecurity, and fostering international cooperation. Considering U.S. foreign policy, cyber diplomacy involves strategic interactions that aim to influence global digital policy, promote security norms, and manage the complexities of cyber conflict and cooperation with other nations.⁷

Automation in cyber diplomacy encompasses using technology to handle repetitive and data-intensive tasks, previously requiring substantial human effort. This includes collecting, sorting, and analyzing vast data amounts related to global cyber threats and opportunities, allowing diplomats and analysts to focus more on strategy and decision-making.⁸ Automation enhances the speed and accuracy of responses to international cyber incidents and policy development.⁹

AI in cyber diplomacy includes using machine learning, natural language processing, and other AI techniques to analyze complex datasets and make predictions. AI is crucial for predictive analytics, processing large data volumes, and strategic planning, enhancing the effectiveness and precision of diplomatic engagements.[10]

---

⁶ Marta Konovalova, "AI AND DIPLOMACY: CHALLENGES AND OPPORTUNITIES," *Journal of Liberty and International Affairs* 9, no. 2 (2023): 699–715, 527, https://doi.org/10.47305/JLIA2392699k.

⁷ A. Georgescu, "Cyber Diplomacy in the Governance of Emerging AI Technologies-A Transatlantic Example," *International Journal of Cyber Diplomacy*, 2022, 15, https://ijcd.ici.ro/documents/3/2022_article_2.pdf.

⁸ Marta Konovalova, "AI AND DIPLOMACY: CHALLENGES AND OPPORTUNITIES," *Journal of Liberty and International Affairs* 9, no. 2 (2023): 699–715, 527, https://doi.org/10.47305/JLIA2392699k.

⁹ Julia Rosinski, Letho Karl, and Russell Eugene, "AI and National Security: The Geopolitical Implications of Autonomous Weapons and Cyber Security" (2023), 3, *OSF*, https://doi.org/10.31219/osf.io/hecxt.

Collectively, these concepts underpin the modernization of U.S. cyber diplomacy efforts, enabling more proactive, strategic, and timely responses in the interconnected global landscape.[11]

Historical Context of U.S. Cyber Diplomacy

The evolution of U.S. cyber diplomacy has been marked by increasing complexity and urgency in addressing global digital threats and opportunities. Emerging from the late 20th century with the advent of the internet, U.S. efforts in cyber diplomacy initially focused on promoting open communication and fostering international collaboration in digital spaces. However, the early 2000s highlighted significant vulnerabilities, as cyber-attacks began targeting national infrastructure and governmental networks. These incidents revealed the limitations of existing diplomatic measures, which were ill-equipped to respond swiftly and effectively to such rapid technological exploits.[12]

This lack of preparedness necessitated a reevaluation of strategies and the integration of more sophisticated technological tools. The U.S. government recognized the need to innovate not just in defensive cyber operations but also in how diplomatic engagements could be conducted in the digital domain. This period catalyzed the integration of technological advancements, including automation and AI, into the diplomatic toolkit, aiming to enhance the nation's capability to anticipate, respond, and negotiate in the fast-paced field of international cyber politics.[13] These challenges set the stage for a transformative shift towards a more proactive and technologically integrated approach in U.S. cyber diplomacy.

---

[10] Petar Radanliev, "Cyber Diplomacy: Defining the Opportunities for Cybersecurity and Risks from Artificial Intelligence, IoT, Blockchains, and Quantum Computing," *Journal of Cyber Security Technology*, (2024): 1-51, 18, doi:10.1080/23742917.2024.2312671.

[11] Philip McDonagh et al., "Knowing What We Ought to Know: The Issues That Face 21st-Century Diplomacy," in *On the Significance of Religion for Global Diplomacy*, 1st ed., vol. 1 (United Kingdom: Routledge, 2021), 38–57, 49, https://doi.org/10.4324/9781003053842-3.

[12] Amel Attatfa, Karen Renaud, and Stefano De Paoli, "Cyber Diplomacy: A Systematic Literature Review," *Procedia Computer Science* 176 (2020): 60-69, 61, https://doi.org/10.1016/j.procs.2020.08.007.

Impact of Automation on Cyber Diplomacy

Automation has fundamentally altered the operational landscape of U.S. cyber diplomacy, enhancing the efficiency and effectiveness of diplomatic communications and data management. By automating repetitive and data-intensive tasks, diplomats can now focus on strategic decision-making and complex negotiations. The automation tools range from simple data collection systems to sophisticated algorithms that analyze and predict diplomatic outcomes based on historical data.[14]

One of the key areas where automation has made a significant impact is in the management of diplomatic communications. Automated systems facilitate the rapid sorting and prioritization of incoming communications, ensuring that critical information reaches decision-makers swiftly. This capability is crucial in crisis situations where timely responses can mitigate potential conflicts or escalate diplomatic engagement as needed.[15]

In terms of data management, automation helps in aggregating and synthesizing vast amounts of data from various sources, including satellite imagery, threat feeds, and international media outlets. This integration provides a comprehensive overview of the global cyber landscape, aiding diplomats in understanding potential threats and opportunities. Automated threat detection systems have been employed to monitor and respond to cyber incidents in real-time, providing U.S. diplomats with actionable intelligence to discuss with international partners.[16]

---

[13] Simon Handler, "The 5×5—the Future of Cyber Diplomacy," *Atlantic Council*, September 29, 2021, 2, https://www.atlanticcouncil.org/commentary/the-5x5-the-future-of-cyber-diplomacy/.

[14] L. Kello, "Digital Diplomacy and Cyber Defence," in *The Oxford Handbook of Digital Diplomacy*, 2024, 513, https://books.google.com/books?hl=en&lr=&id=1bTnEAAAQBAJ&oi=fnd&pg=PA121&dq=automation+U.S.+cyber+diplomacy&ots=K0QNnOotbB&sig=ByrCoaKfQcrRk9JWKrunitffg9g.

[15] L. Suchman, "Imaginaries of omniscience: Automating intelligence in the US Department of Defense," *Social Studies of Science* (2023),768, https://journals.sagepub.com/doi/abs/10.1177/03063127221104938.

These automated tools not only improve operational efficiencies but also enhance the accuracy and reliability of the data on which diplomatic decisions are based. As cyber diplomacy requires handling increasingly complex and sensitive information, the role of automation becomes even more critical, ensuring that the U.S. maintains its leadership and strategic advantage in the global digital domain.[17]

Strategic Benefits of AI in Cyber Diplomacy

AI profoundly enhances the strategic dimensions of U.S. cyber diplomacy, providing essential tools that streamline complex decision-making processes, bolster threat intelligence, and refine diplomatic strategies in real-time. AI's ability to process and analyze large volumes of data rapidly translates into a more nuanced understanding of global cyber trends and potential threats, allowing U.S. diplomats to anticipate challenges and seize opportunities in the digital domain.[18]

AI excels in predictive analytics, which involves using algorithms to parse through extensive data sets, from cybersecurity alerts to global political shifts, to forecast potential security threats and geopolitical events. This capability enables the U.S. to proactively adjust its cyber diplomacy strategies, ensuring readiness and response before potential crises materialize.

---

[16] D.T. Varela, "Navigating Cyber Diplomacy in the Governance of Emerging AI Technologies: Lessons from Transatlantic Cooperation," *Journal of Artificial Intelligence General Science*, 2024, 111, http://jaigs.org/index.php/JAIGS/article/view/29.

[17] A. Georgescu, "Cyber Diplomacy in the Governance of Emerging AI Technologies," *International Journal of Cyber Diplomacy*, 2022, 13.

[18] Alexandra-Cristina Dinu, "Cyber Diplomacy and Artificial Intelligence," *Proceedings of the International Conference on Cybersecurity and Cybercrime - 2023*, 90.

For instance, AI-driven systems are employed to monitor international cyber activities, alerting diplomats to anomalous behaviors that could indicate imminent threats or breaches.[19]

AI also significantly contributes to strategic planning by offering simulations of diplomatic scenarios based on vast historical and current data. These simulations assist policymakers in understanding possible outcomes and developing contingency plans. This strategic advantage was evident when AI models provided U.S. negotiators with insights during international cyber treaty discussions, enabling them to advocate effectively for terms that enhance global cyber stability and align with U.S. security interests.[20]

AI systems support real-time decision-making in cyber diplomacy by integrating dynamic data streams, such as live cyber-attack maps and diplomatic communications, which are vital during high-stakes negotiations or cyber incidents. This integration allows diplomats to make informed decisions swiftly, a critical capability in fast-paced environments where delays can escalate vulnerabilities.[21]

During a multinational effort to counter cyber espionage, AI tools analyzed data breach patterns and identified the probable sources and methods used. This critical information guided diplomatic dialogues, leading to a coordinated response that effectively mitigated the espionage activities.[22] Subsequently, AI algorithms played a key role in identifying and countering misinformation campaigns during election cycles in several countries, thereby supporting global democratic integrity and reducing tensions.[23]

---

[19] R. Williams and L. Otto, "Artificial intelligence as a tool of public diplomacy: Communication between the United States and Iran," *The Thinker*, 2022, 37, https://journals.uj.ac.za/index.php/The_Thinker/article/download/1171/750.
[20] Marta Konovalova, "AI AND DIPLOMACY: CHALLENGES AND OPPORTUNITIES," *Journal of Liberty and International Affairs* 9, no. 2 (2023): 525.
[21] GNC Cucinelli, "Cybersecurity and the Risk of Artificial Intelligence," *ResearchGate*, 2022, 109, https://www.researchgate.net/profile/Agnes-Santha/publication/374290230_Qualitative_Methodologies/links/6517083b3ab6cb4ec6a94389/Qualitative-Methodologies.pdf#page=88.
[22] Petar Radanliev, "Cyber Diplomacy: Defining the Opportunities," *Journal of Cyber Security Technology*, (2024): 9.

Through these strategic applications, AI not only supports but advances the U.S.'s capabilities in cyber diplomacy. It transforms traditional diplomatic methods into more dynamic, precise, and effective processes, underscoring the critical role AI plays in maintaining national security and fostering international cooperation in the digital domain.[24] This strategic integration of AI ensures that U.S. diplomacy can navigate the complexities of the digital domain with confidence and foresight.

Addressing Challenges and Risks with AI and Automation

While automation and AI significantly enhance the capabilities of U.S. cyber diplomacy, their integration into sensitive diplomatic spheres brings potential risks and challenges that must be managed carefully. These issues primarily include ethical dilemmas, security vulnerabilities, and the risk of over-reliance on technology which can potentially undermine human judgment.

The deployment of AI in cyber diplomacy raises concerns regarding privacy, surveillance, and the potential for biased decision-making. AI systems can sometimes operate as "black boxes," where the decision-making process is opaque.[25] This lack of transparency can lead to decisions that may inadvertently reflect embedded biases in the data or algorithms, affecting international relations negatively. Ensuring ethical guidelines that promote transparency and fairness is critical in mitigating these risks.[26]

---

[23] R. Williams and L. Otto, "Artificial intelligence as a tool of public diplomacy," *The Thinker*, 2022, 35.

[24] F. Roumate, "Artificial Intelligence and Digital Diplomacy," in *Artificial Intelligence To Strengthen or to Replace Traditional Diplomacy*, ed. Marius Vacarelu (Springer, 2022), 20, https://www.researchgate.net/profile/Marius-Vacarelu/publication/354267165_Artificial_Intelligence_To_Strengthen_or_to_Replace_Traditional_Diplomacy/links/64070a5f0cf1030a567d9d9a/Artificial-Intelligence-To-Strengthen-or-to-Replace-Traditional-Diplomacy.pdf.

[25] Petar Radanliev, "Cyber Diplomacy: Defining the Opportunities," *Journal of Cyber Security Technology*, (2024): 5.

AI and automated systems, by their nature, are susceptible to various forms of cyber threats including data breaches, and manipulation of algorithmic decision-making through tainted data inputs. Such vulnerabilities could potentially be exploited to influence diplomatic actions or to mislead policymakers. Implementing robust cybersecurity measures and continuously updating them against emerging threats is essential for safeguarding these technologies.[27]

To counter the risk of over-reliance on automation and AI, it is crucial to maintain strong human oversight. While AI can process and analyze data at unprecedented speeds and scale, human judgment is indispensable, especially in complex diplomatic negotiations where nuances and soft skills play a pivotal role. Establishing protocols that ensure decisions are reviewed and ratified by human agents can help balance the benefits of AI with the need for human expertise and ethical considerations.[28]

To address these challenges, developing comprehensive AI governance frameworks is necessary. These should include ethical standards for AI deployment in diplomacy, rigorous security protocols to protect data integrity, and training programs for diplomats to effectively use and oversee AI technologies. Additionally, international cooperation to set global norms and

---

[26] F. Roumate, "Artificial Intelligence and Digital Diplomacy," in *Artificial Intelligence To Strengthen or to Replace Traditional Diplomacy*, 2022, 20.
[27] GNC Cucinelli, "Cybersecurity and the Risk of Artificial Intelligence," *ResearchGate*, 2022, 104.
[28] Marta Konovalova, "AI AND DIPLOMACY: CHALLENGES AND OPPORTUNITIES," *Journal of Liberty and International Affairs* 9, no. 2 (2023): 525.

standards for the use of AI in international relations can help mitigate risks and harmonize efforts.[29]

By implementing these solutions, the U.S. can enhance the strategic advantages of AI and automation in cyber diplomacy while minimizing potential downsides. This balanced approach will ensure that these technologies serve as tools for advancing national interests and global stability, rather than becoming sources of unintended ethical and security issues.

Conclusion

In this paper, the transformative impact of automation and artificial intelligence (AI) on enhancing U.S. cyber diplomacy has been thoroughly examined. These technological advancements have proven to be pivotal in managing the complexity and speed required in modern diplomatic engagements, enabling more informed decision-making, increased efficiency, and fortified security.[30]

Automation has significantly improved operational efficiencies in handling diplomatic communications and data management, allowing for a more streamlined approach to the ever-growing volumes of data that diplomats must navigate. So far AI has extended the strategic capabilities of cyber diplomacy through threat intelligence platforms, providing diplomats with tools that consolidate threat feeds.

However, the integration of these technologies is not without challenges. Ethical dilemmas, security vulnerabilities, and the potential for over-reliance on automated systems pose substantial risks. Addressing these issues requires robust frameworks for AI governance, including ethical guidelines, stringent security measures, and the preservation of human oversight within AI-driven processes.

---

[29] A. Georgescu, "Cyber Diplomacy in the Governance of Emerging AI Technologies," *International Journal of Cyber Diplomacy*, 2022, 21.

[30] Simon Handler, "The 5×5—the Future of Cyber Diplomacy," *Atlantic Council*, September 29, 2021, 2.

By effectively mitigating these risks and harnessing the potential of automation and AI, U.S. cyber diplomacy can transcend traditional limitations, paving the way for a new era of diplomatic practice that is not only reactive but also proactive and preemptive. As the digital domain and cyberspace continue to evolve, our approaches to international relations and security must also adapt accordingly. Automation and AI stand at the forefront of this evolution, reshaping the way the U.S. interacts on the global stage and safeguarding its interests in international cyber diplomacy.